\documentclass[preprint,12pt]{elsarticle}

\usepackage{silence}
\usepackage{placeins}
\usepackage{amssymb}
\usepackage{amsmath}
\usepackage{graphicx}
\usepackage{booktabs}
\usepackage{amsmath}
\usepackage{enumitem}
\usepackage{caption}
\usepackage{subcaption}
\usepackage{hyperref}
\usepackage{amsthm}
\usepackage[final]{microtype}
\journal{Journal of Informatics}

\begin{document}

\begin{frontmatter}

\title{Mapping the Climate–Health Evidence Base (2007–2023): A Bibliometric, Statistical, and NLP Multi-Label Text Analysis of 22,695 Records}

\author[aff1]{Dhruv Dixit}
\author[aff2,aff3]{Janine Molino}

\affiliation[aff1]{organization={Stevens Institute of Technology},
            city={Hoboken},
            postcode={07030},
            state={New Jersey},
            country={United States}}
\affiliation[aff2]{organization={Brown University},
            city={Providence},
            postcode={02912},
            state={Rhode Island},
            country={United States}}
\affiliation[aff3]{organization={Rhode Island Hospital},
            city={Providence},
            postcode={02912},
            state={Rhode Island},
            country={United States}}

\begin{abstract}
We analyzed a curated climate--health bibliographic corpus of 22{,}695 multi-labeled records from 2007--2023 to characterize growth, thematic concentration, and evolving methods. Annual publication counts rose sharply, with multiple change-points indicating phase-structured expansion; Negative Binomial models estimated roughly 10--11\% year-over-year growth. Exposure--health co-occurrence departed strongly from independence, with canonical hazard--outcome dyads (e.g., extreme heat with heat-related impacts; floods/hurricanes with mental health) occurring far more often than expected even after accounting for marginal term popularity. A hierarchical logistic model for asthma-tagged records showed strong alignment with air-pollution-related exposures (including ozone and particulate matter) and relative under-representation of generic heat/temperature terms. Methodologically, modeling timescales shifted toward longer horizons over time, while at least one legacy method tag declined. Finally, we detected time- and geography-dependent annotation completeness, including decreased exposure-term coding in recent years, underscoring the need to model missingness when interpreting temporal trends.
\end{abstract}


\end{frontmatter}

\section{Introduction}
Climate change was increasingly recognized as a cross-cutting determinant of health, yet the evidence base was dispersed across hazards (e.g., heat, floods, air pollution), outcomes (e.g., respiratory, cardiovascular, mental health), geographies, and methodological approaches. To characterize how this literature has evolved---and where it concentrates attention---we analyzed a curated bibliographic corpus\footnote{Dataset available at: \url{https://www.datalumos.org/datalumos/project/222402/version/V1/view?path=/datalumos/222402/fcr:versions/V1}} of 22{,}695 climate--health records spanning 2007--2023 with rich controlled-vocabulary, multi-label annotations for exposures, health impacts, geography, and methods.\footnote{Anonymized Code: \url{https://github.com/my-anonymous-account-4-papers/anon-informetrics-2026}}

In this study, we aim to (i) quantify temporal growth and structural shifts in publication volume, (ii) identify which exposure--health relationships are disproportionately emphasized relative to baseline marginals, (iii) evaluate predictors of a representative respiratory subtopic (asthma), (iv) assess shifts in modeling timescales/methods over time, and (v) diagnose time- and geography-dependent coding completeness that can bias trend inference.

\section{Background}
\subsection{Climate Change and Human Health: Scope and Evolution of the Field}

Climate change has emerged as one of the most significant global challenges of the 21st century, with far-reaching implications that extend beyond environmental systems into human health and public well-being. Over the past several decades, the scientific community has increasingly recognized that climate-related phenomena—such as rising temperatures, extreme weather events, and shifting ecological patterns—are closely intertwined with population-level health outcomes \cite{Verner2016}.
\\\\
Early research in this field was relatively sparse and fragmented, but since the early 2000s, studies examining the relationship between climate change and human health have grown steadily in both volume and scope. Bibliometric evidence shows that this growth accelerated markedly after 2007, reflecting increased global awareness, improved data availability, and stronger interdisciplinary collaboration \cite{Verner2016}. More recent analyses confirm that research output continued to expand substantially between 2012 and 2021, indicating sustained and growing scientific interest in climate--health interactions \cite{Bartlett2024}.
\\\\
As the field has matured, research attention has concentrated on a set of dominant climate exposures and health outcomes. Among these, heat-related risks have consistently emerged as a central focus, given their direct association with excess mortality, cardiovascular strain, and heat-related illness, particularly among vulnerable populations \cite{Klingelhofer2023}. In parallel, infectious diseases have received considerable attention due to their sensitivity to climatic variables such as temperature, precipitation, and vector ecology \cite{Li2020,Sweileh2020}. These thematic concentrations suggest that the climate--health literature is not evenly distributed across health domains, but rather clustered around specific exposure--outcome pathways.

\subsection{Structure and Trends in the Climate--Health Research Literature}

Beyond thematic focus, several studies have examined how knowledge in this field is structured and disseminated. Large-scale bibliometric and scientometric analyses demonstrate that climate--health research exhibits distinct collaboration networks, geographic imbalances, and evolving research priorities. For example, knowledge-mapping and network-based approaches have revealed that the majority of highly cited work originates from high-income countries, while regions most vulnerable to climate impacts remain comparatively underrepresented in the literature \cite{Zhao2020}. Citation-based analyses further suggest that certain health risks—particularly infectious diseases—receive disproportionate attention relative to other outcomes such as mental health or chronic disease \cite{Kolsky2023}.
\\\\
More recently, comprehensive bibliometric studies have begun to explicitly frame climate change and public health research as a dynamic, evolving system rather than a collection of isolated studies. These analyses identify clusters of research activity, track shifts in dominant themes over time, and highlight persistent gaps in coverage and methodological approaches \cite{Raval2024}. Together, this body of work underscores the value of quantitative, large-scale analytical methods for synthesizing an increasingly complex and rapidly expanding literature.
\\\\
Taken together, these findings suggest that while the climate--health literature has expanded rapidly and addressed a wide range of exposures and outcomes, it is characterized by substantial heterogeneity in both thematic focus and analytical approaches. This complexity motivates the use of large-scale quantitative methods to systematically characterize research trends, classify commonly used statistical analyses, and identify gaps that may inform future investigation.

\subsection{Gaps in the Existing Literature}
Despite the rapid increase in climate-health research, significant gaps remained in our understanding of the field's structural evolution. While recent bibliometric analyses successfully quantified the volume of growth they often relied on descriptive statistics such as simple publication counts or country-level rankings, rather than rigorous statistical modeling. Previous overviews have largely relied on descriptive mapping techniques, such as calculating the frequency of specific climate impact with health outcome combinations \cite{rocque2021health}. This overview of systematic reviews was conducted by Rocque and colleagues, led by Rhea J. Rocque (Prairie Climate Centre, University of Winnipeg), with coauthors based at Université Laval’s Faculty of Medicine and VITAM Research Centre for Sustainable Health, plus collaborators from CHUQ Research Centre and the University of Toronto (Li Ka Shing Knowledge Institute and Dalla Lana School of Public Health). In their study, the authors explicitly noted that due to the heterogeneity of the included data, they were unable to conduct meta-meta-analyses or rigorous statistical syntheses. This adds limitations to infer the results with a statistically comprehensive model. 

\subsection{Rationale for Using a Multi-Method Statistical Approach}
The climate--health evidence base was inherently multi-modal: each reference contained near-complete bibliographic metadata (e.g., year, title, reference type) and, for many records, unstructured text (abstracts), while substantive content is additionally captured through curated multi-label taxonomies (e.g., exposure terms, health impact terms, special topic tags). These components varied in completeness across the dataset (Figure~\ref{fig:completeness_schema}), and they encoded different scientific signals (publication volume and venues vs.\ topical framing vs.\ co-occurrence structure). As a result, no single statistical model could faithfully answer all study questions without either discarding large portions of the dataset or over-interpreting sparsely annotated fields.

We therefore used a multi-method strategy in which each method targets a distinct inferential goal while remaining consistent with the structure and coverage of the underlying data. First, we used descriptive and bibliometric analyses to characterize the volume of evidence over time and across geographies using the full corpus where possible. Second, to quantify relationships between coded climate exposures and coded health impacts, we used multi-hot representations of the taxonomy fields and fit regression models (including hierarchical specifications when appropriate) that directly map interpretable term indicators to outcomes of interest. This provided effect estimates that were straightforward to communicate and align with the curated ontology.

Third, we acknowledged that document-level covariate topic models---such as Structural Topic Models (STM)---were well suited for estimating topic prevalence as a function of metadata. However, our primary objective was to obtain a stable, interpretable topic basis over a large corpus with minimal additional modeling assumptions and computational overhead. We therefore used an LDA-style topic model to infer latent themes from titles/abstracts and then examined topic prevalence and trends using downstream regression against time, geography, and other covariates. This two-stage approach preserves the key advantage of covariate-linked interpretation while improving robustness and reproducibility for large-scale exploratory mapping.

Finally, the exposure--health co-occurrence structure can be viewed through a network lens. Although Exponential Random Graph Models (ERGMs) provide a flexible framework for modeling dependence in networks, they are often computationally challenging and can be unstable for large, sparse bipartite graphs with many possible edges (as in high-dimensional term co-occurrence networks). Instead, we modeled association patterns using scalable count-based/log-linear formulations (with degree/volume adjustments where relevant) and complementary permutation/robustness checks. This preserves the core objective of identifying enriched exposure--health pairings while remaining tractable and transparent at the scale of the evidence base.

Overall, this multi-method design ensures that (i) broad bibliometric conclusions leverage maximal coverage, (ii) term-level inferences remain interpretable within the ontology, and (iii) text-derived themes and co-occurrence structure are incorporated in a way that is computationally stable and easy to audit. STM and ERGM-based extensions remained promising directions for future work when the analytic goal requires fully joint estimation of covariate-linked topic structure or explicit dependence modeling in networks.

\section{Hypothesis}
We operationalized hypotheses aligned to the project analysis plan:
\begin{itemize}[leftmargin=3em, label=\scriptsize\textbullet, itemsep=0pt, parsep=0pt, topsep=2pt, partopsep=0pt]
    \item \textbf{H1 (Trend):} The climate--health literature increased substantially over 2007--2023, with identifiable change-points (waves of acceleration/plateau).
    \item \textbf{H2 (Association structure):} Specific exposure--health dyads (canonical hazard--outcome pairings) occured far more often than expected under independence.
    \item \textbf{H3 (Geography/coding):} Geographic strata differed in topic emphasis and/or coding completeness, implying potential bias if not modeled.
    \item \textbf{H4 (Methods/timescales):} Modeling timescales shifted toward longer horizons over time; some legacy method tags declined.
    \item \textbf{H5 (Venue dynamics):} Growth is not only due to more venues; within-journal climate--health output increased over time.
    \item \textbf{H6 (Topic heterogeneity):} Topic prevalence varied by time, geography, and reference type.
\end{itemize}

\section{Data Management/Manipulation}
\subsection{Data source and structure}
The dataset\footnote{Dataset link: \url{https://www.datalumos.org/datalumos/project/222402/version/V1/view?path=/datalumos/222402/fcr:versions/V1}} was obtained via a targeted Google search for publicly available datasets related to climate change literature. It was assembled for the purpose of analyzing trends in topics, methods, and the direction of research activity in the climate literature domain. 

The dataset contained 22{,}695 records across 29 columns (2007--2023). Records were predominantly journal articles (\(\sim 97.6\%\)) with metadata (title, journal, year, reference type) and multiple pipe-separated controlled-vocabulary fields for exposures, health impacts, geography, methods, and special topics.

\subsection{Cleaning and canonicalization}
We standardized core categorical fields (e.g., geography codes, reference type) and treated invalid sentinel values (e.g., year \(=0\)) as missing. Term strings were trimmed/normalized to support consistent multi-label parsing.

\subsection{Multi-label expansion and design-matrix creation}
Pipe-separated multi-label fields were parsed into lists and expanded into multi-hot indicator matrices (binary presence/absence). To manage sparsity, frequency thresholds were used when forming modeling-ready predictors. A unified modeling table combined metadata (year, journal, geography codes, reference type) with multi-hot term families using consistent prefixes (e.g., \texttt{exposure\_\_*\ }, \texttt{health\_impact\_\_*\ }).

\subsection{Missingness and bias variables}
We created indicators for whether a record contained an abstract, exposure terms, and health-impact terms to support missingness diagnostics and sensitivity analyses.

\subsection{Exports}
Derived artifacts (multi-hot matrices, final design matrix, yearly count tables, and top-term summaries) were exported to support reproducibility and reporting.

\section{Statistical Analysis Plan and Justification}

\subsection{Bibliometric growth models and structural change}
Let $Y_t$ denote the number of records published in year $t \in \{2007,\dots,2023\}$. We first reported descriptive bibliometrics (annual totals, reference-type composition, and geography-code distributions), then modeled the long-run time trend using count regression. Specifically, we fit Poisson and Negative Binomial (NB) generalized linear models (GLMs),
\begin{equation}
Y_t \sim \mathrm{Poisson}(\mu_t), 
\qquad \log(\mu_t) = \alpha + \beta (t-\bar t),
\end{equation}
and, to accommodate overdispersion common in bibliometric series,
\begin{equation}
Y_t \sim \mathrm{NB}(\mu_t, k), 
\qquad \mathrm{Var}(Y_t)=\mu_t + \frac{\mu_t^2}{k},
\qquad \log(\mu_t) = \alpha + \beta (t-\bar t),
\end{equation}
where $\exp(\beta)$ was interpreted as the multiplicative change in expected publications per year (incidence rate ratio, IRR). Model preference (Poisson vs.\ NB) was justified by dispersion diagnostics (e.g., Pearson $\chi^2$/df, residual patterns) and information criteria (AIC/BIC).

To identify non-smooth growth, we estimated change-points in $\{Y_t\}$ using (i) a joinpoint-style segmentation via PELT (Pruned Exact Linear Time), selecting breakpoints by minimizing a penalized cost, and (ii) a Bayesian Poisson change-point model with unknown break year $\tau$:
\begin{equation}
Y_t \sim \mathrm{Poisson}(\mu_t),\quad
\log(\mu_t) =
\begin{cases}
\alpha_1 + \beta_1 (t-\bar t), & t \le \tau,\\
\alpha_2 + \beta_2 (t-\bar t), & t > \tau,
\end{cases}
\end{equation}
with a discrete prior on $\tau$ over the observed year range. These complementary approaches provided both point estimates (PELT) and uncertainty-aware inference (Bayesian) for structural shifts.

\subsection{Exposure--health structure via high-dimensional contingency screening}
For records $n=1,\dots,N$, let $X_{ni}\in\{0,1\}$ indicated whether exposure term $i$ is present and $H_{nj}\in\{0,1\}$ indicated whether health-impact term $j$ is present after multi-label expansion and multi-hot encoding. We defined the exposure$\times$health co-occurrence matrix
\begin{equation}
C_{ij} = \sum_{n=1}^N X_{ni} H_{nj},
\end{equation}
and screen for over-/under-represented dyads under an independence (log-linear) baseline. Under independence, the expected count was
\begin{equation}
E_{ij} = \frac{C_{i+}\,C_{+j}}{C_{++}},
\end{equation}
where $C_{i+}=\sum_j C_{ij}$, $C_{+j}=\sum_i C_{ij}$, and $C_{++}=\sum_{i,j} C_{ij}$. We computed standardized residuals
\begin{equation}
z_{ij}=\frac{C_{ij}-E_{ij}}{\sqrt{E_{ij}}},
\end{equation}
and convert $z_{ij}$ to two-sided $p$-values using a Normal approximation to flag dyads with unusually high or low co-occurrence. Because the number of tested dyads is large, we control the false discovery rate (FDR) using the Benjamini--Hochberg (BH) procedure: if $p_{(1)}\le \cdots \le p_{(m)}$ are sorted $p$-values, we declare significance for all $p_{(i)}\le (i/m)q$, where $q$ is the target FDR level. Significant dyads are summarized via ranked tables and heatmaps of $z_{ij}$ (or observed/expected ratios).

\subsection{Network-style edge models and record-level regression}
To distinguish ``popular terms'' from ``specific pairings,'' we model the positive exposure--health edges as a bipartite weighted network with edge weight $C_{ij}$. We fit a Poisson edge model with degree-based offsets to account for overall term prevalence. Let $d_i=C_{i+}$ and $s_j=C_{+j}$; an intercept-only offset model is
\begin{equation}
C_{ij} \sim \mathrm{Poisson}(\lambda_{ij}), 
\qquad \log(\lambda_{ij}) = \alpha + \log d_i + \log s_j,
\end{equation}
and we use residuals and FDR-adjusted tests to identify dyads that remain unusually strong/weak after controlling for marginal popularity.

We summarize departures from the offset baseline using Pearson residuals
\begin{equation}
r_{ij}=\frac{C_{ij}-\hat{\lambda}_{ij}}{\sqrt{\hat{\lambda}_{ij}}}.
\end{equation}

For record-level inference on a representative binary outcome (e.g., whether a record is tagged with an asthma health-impact term), we fit logistic regression models using exposure indicators and core metadata (year, reference type, geography codes) as predictors:
\begin{equation}
Y_n \sim \mathrm{Bernoulli}(p_n),\qquad 
\mathrm{logit}(p_n) = \gamma_0 + \gamma_1 (t_n-\bar t) + \mathbf{x}_n^\top \boldsymbol{\gamma} + \mathbf{w}_n^\top \boldsymbol{\delta},
\end{equation}
where $\mathbf{x}_n$ are exposure indicators and $\mathbf{w}_n$ are metadata covariates (with categorical factors expanded to dummies). Because records cluster within journals (and potentially geography strata), we report cluster-robust standard errors where appropriate. To mitigate instability from sparsity or quasi-separation, we employ regularization (e.g., L1/L2) or move to partial pooling.

We therefore also fit hierarchical logistic models with random intercepts for geography and journal:
\begin{equation}
\begin{aligned}
\mathrm{logit}(p_n) &= \beta_0 + \beta_1 (t_n-\bar t) + \mathbf{x}_n^\top \boldsymbol{\beta}
+ u_{g[n]} + v_{j[n]}, \\
u_g &\sim \mathcal{N}(0,\sigma_g^2), \qquad
v_j \sim \mathcal{N}(0,\sigma_j^2).
\end{aligned}
\end{equation}
which stabilizes estimation and yields interpretable variance components for between-geo and between-journal heterogeneity. When modeling multiple common health outcomes jointly, we use multivariate logistic formulations (e.g., latent-factor or correlated random-effect structures) to capture residual outcome correlation beyond shared covariates.

\paragraph{Multivariate probit (latent correlation)}
For $K$ coarse outcomes, let $y_{nk}\in\{0,1\}$ indicate whether record $n$ carries outcome $k$.
We introduce latent variables $\mathbf{z}_n\in\mathbb{R}^K$:
\begin{align}
\mathbf{z}_n &\sim \mathcal{N}\!\left(\mathbf{X}_n\mathbf{B},\,\mathbf{\Sigma}\right), \\
y_{nk} &= \mathbb{I}(z_{nk} > 0),
\end{align}
where $\mathbf{\Sigma}$ induces residual dependence between outcomes after conditioning on covariates.
We summarize dependence via the latent correlation matrix
$\mathbf{R}=\mathrm{diag}(\mathbf{\Sigma})^{-1/2}\mathbf{\Sigma}\,\mathrm{diag}(\mathbf{\Sigma})^{-1/2}$.

\subsection{Topic models, methods/timescales, and robustness to missingness}
To characterize latent themes in titles/abstracts, we fit LDA on preprocessed text (tokenization, stopword removal, vocabulary filtering) and obtain per-document topic proportions $\theta_{nk}$. We then assess topic heterogeneity by regressing transformed topic prevalence on covariates such as year, geography, and reference type (e.g., $\mathrm{logit}(\theta_{nk})$ as the response in linear models), emphasizing uncertainty and multiple-comparison considerations across topics.

For sparse methodological fields, we model \texttt{model\_timescale} as an ordered outcome using ordered logistic regression:
\begin{equation}
\Pr(Y_n \le r \mid \eta_n) = \mathrm{logit}^{-1}(\kappa_r - \eta_n),
\qquad 
\eta_n = \alpha + \beta (t_n-\bar t),
\end{equation}
and analyze rare method indicators with penalized logistic regression or exact tests (e.g., Fisher's exact test) when small cell counts preclude asymptotic inference.

Finally, we quantify time- and geography-dependent coding completeness by modeling missingness indicators (e.g., \texttt{has\_abstract}, \texttt{has\_exposure\_terms}, \texttt{has\_health\_terms}) as logistic functions of year and metadata. Because incomplete coding can confound temporal trends, key findings are subjected to sensitivity analyses (e.g., alternative term-frequency thresholds for design-matrix construction and re-fitting core models). Where feasible and defensible, multiple imputation is used for a limited set of covariates; however, given substantial structural missingness in some sparse controlled-vocabulary fields, imputation is treated as exploratory and accompanied by diagnostic checks.

\begin{figure}[t]
    \centering
    \begin{minipage}[t]{0.48\linewidth}
        \centering
        \includegraphics[width=\linewidth]{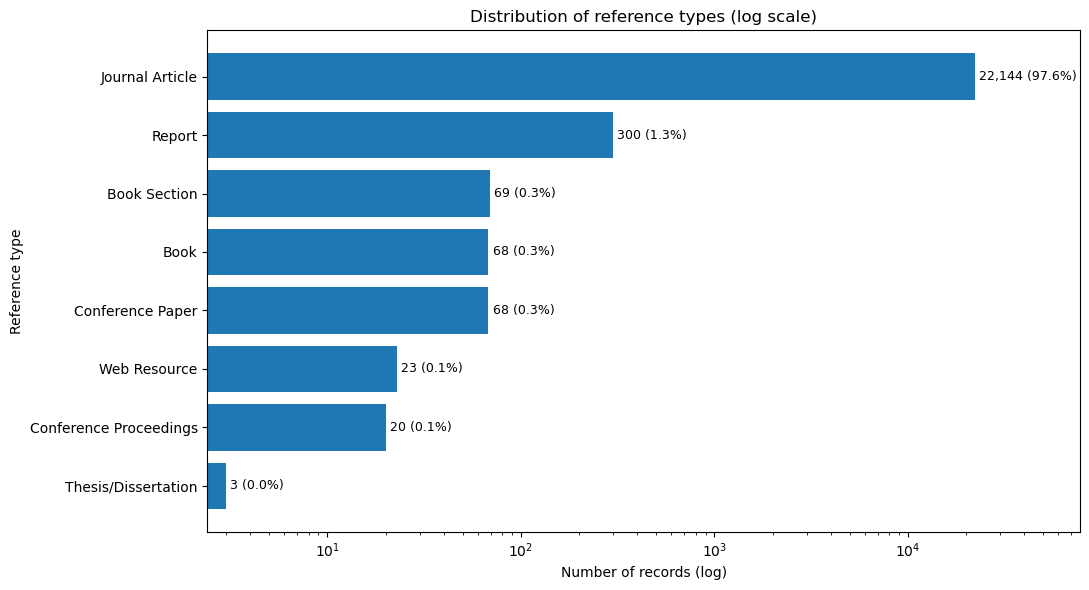}
        \subcaption{}
        \label{fig:ref-type}
    \end{minipage}\hfill
    \begin{minipage}[t]{0.48\linewidth}
        \centering
        \includegraphics[width=\linewidth]{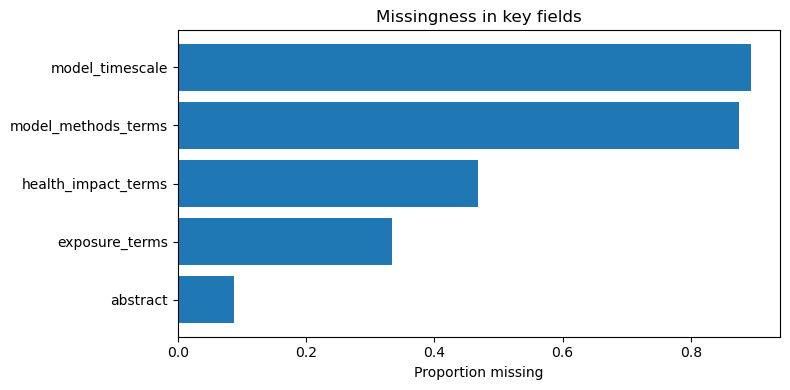}
        \subcaption{}
        \label{fig:missingness}
    \end{minipage}

    \caption{Corpus composition and annotation completeness. (a) Reference-type distribution. (b) Missingness across key annotation fields.}
    \label{fig:corpus-summary-1}
\end{figure}

\section{Results}

\begin{figure}[t]
    \centering
    \begin{minipage}[t]{0.48\linewidth}
        \centering
        \includegraphics[width=\linewidth]{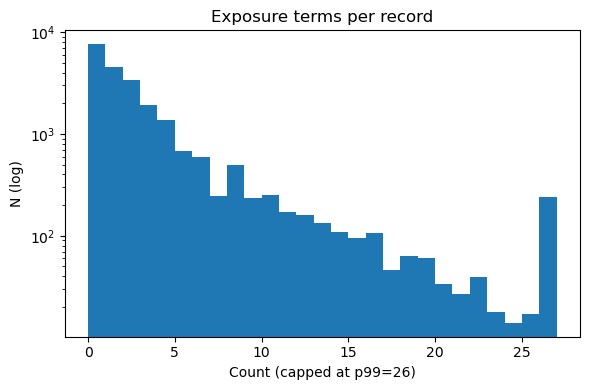}
        \subcaption{}
        \label{fig:exposure-hist}
    \end{minipage}\hfill
    \begin{minipage}[t]{0.48\linewidth}
        \centering
        \includegraphics[width=\linewidth]{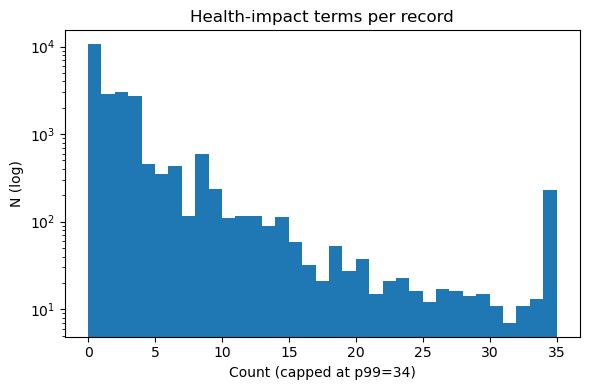}
        \subcaption{}
        \label{fig:health-hist}
    \end{minipage}

    \caption{Corpus composition and annotation completeness. (c) Distribution of exposure-term counts per record. (d) Distribution of health-impact-term counts per record.}
    \label{fig:corpus-summary-2}
\end{figure}

\subsection{Corpus composition, coverage, and temporal/venue structure}
The corpus contains 22{,}695 records and is dominated by journal articles (Figure~\ref{fig:ref-type}). Controlled-vocabulary coverage is sparse: exposure terms appear for a majority but not all records, health-impact terms for roughly half, and method/timescale fields are highly sparse (Figure~\ref{fig:missingness}). Term counts per record are strongly right-skewed, with most records lightly annotated and a long tail of heavily annotated records (Figures~\ref{fig:exposure-hist}--\ref{fig:health-hist}).

\begin{figure}[ht]
    \centering
    \includegraphics[width=0.8\linewidth]{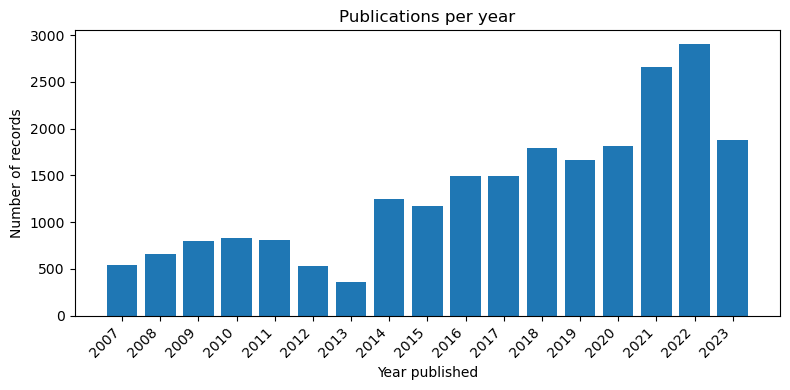} 
    \caption{Publications per year (bar chart) with joinpoint breakpoints (approximately 2011, 2016, 2021, 2023).}
    \label{fig:pubs-joinpoints}
\end{figure}

Publication volume increases sharply from 2007 to the early 2020s (Figure~\ref{fig:pubs-joinpoints}), with joinpoints at approximately 2011, 2016, 2021, and 2023 indicating phase-structured growth. Negative Binomial trend models estimate \(\sim 10\%\) year-over-year growth (IRR \(\approx 1.10\)), and a journal-year Negative Binomial model shows strong within-journal growth (year effect IRR \(\approx 1.11\)).

\begin{figure}[t]
    \centering

    \begin{minipage}[t]{0.49\linewidth}
        \centering
        \includegraphics[width=\linewidth]{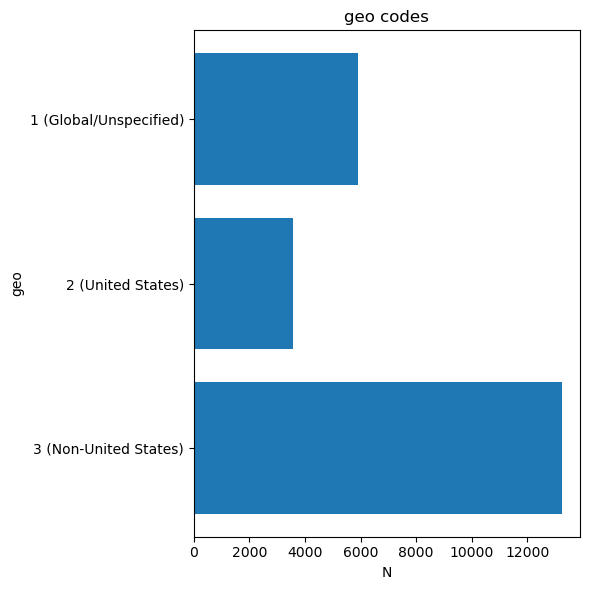}
        \subcaption{}
        \label{fig:geo}
    \end{minipage}\hfill
    \begin{minipage}[t]{0.49\linewidth}
        \centering
        \includegraphics[width=\linewidth]{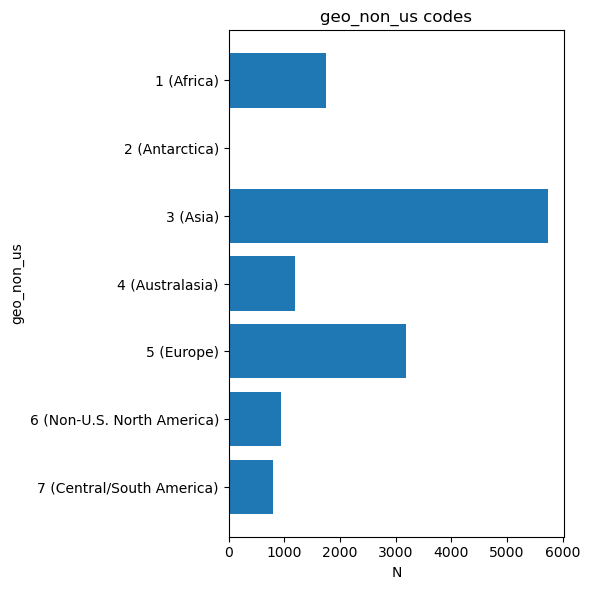}
        \subcaption{}
        \label{fig:geo-nonus}
    \end{minipage}

    \caption{Geographic coding distributions in the corpus. (a) Coarse geographic strata (\texttt{geo}). (b) Non-U.S. geographic strata (\texttt{geo\_non\_us}). X-axis codes in the results section.}
    \label{fig:geo-side-by-side}
\end{figure}

Geography is captured by U.S. geographic strata -- \texttt{geo} (3 levels) and non-U.S. geographic strata -- \texttt{geo\_non\_us} (7 levels). \texttt{geo} is highly imbalanced: code-3 (non-U.S.) accounts for 13{,}233/22{,}695 (58.3\%), followed by code-1 (Global/unspecified) at 5{,}901; 26.0\% and code-2 (U.S.) at 3{,}561; 15.7\% (Figure~\ref{fig:geo}). \texttt{geo\_non\_us} is missing for 9{,}106/22{,}695 records (40.1\%), consistent with conditional applicability (Figure~\ref{fig:geo-nonus}); among non-missing records (n=13{,}589), code-3 (Asia) is most frequent (5{,}738; 42.2\%), followed by code-5 (Europe) at 3{,}184; 23.4\% and code-1 (Africa) at 1{,}741; 12.8\%, while code-2 (Antarctica) is super-rare at just 7 (0.1\%).

\subsection{Dataset characterization and annotation coverage}

Figure~\ref{fig:completeness_schema} summarizes column-wise completeness and highlights that the dataset combines near-complete bibliographic metadata with partially available annotation fields. Core reference descriptors (e.g., reference identifiers, reference type, title text, and publication year) are essentially complete, and abstracts are available for the majority of records (91.3\%). In contrast, ontology-derived content annotations are present for subsets of the corpus: exposure terms are available for 66.5\% of records and health impact terms for 53.2\%. Modeling metadata is substantially sparser, with model methods terms and model timescale available for 12.5\% and 10.6\% of records, respectively. These completeness patterns define the analytic denominators used throughout the Results: bibliometric analyses leverage the full corpus where possible, whereas term-based association and modeling summaries are reported on the relevant non-missing subsets.

\subsection{Substantive patterns: dyad concentration, outcome-specific associations, methods, and missingness}

\begin{figure}[ht]
    \centering
    \includegraphics[width=0.92\linewidth]{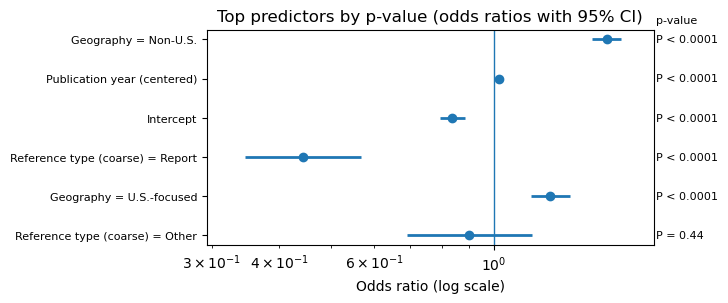}
    \caption{\textbf{Predictors of health-impact annotation availability.} Odds ratios (log scale) with 95\% confidence intervals from a logistic regression predicting whether a record has non-missing \texttt{health\_impact\_terms} as a function of centered publication year, geography, and broad reference type (Journal Article as baseline). Points are ordered by p-value.}
    \label{fig:missingness-logit-health}
\end{figure}

\begin{figure}[ht]
    \centering
    \includegraphics[width=0.92\linewidth]{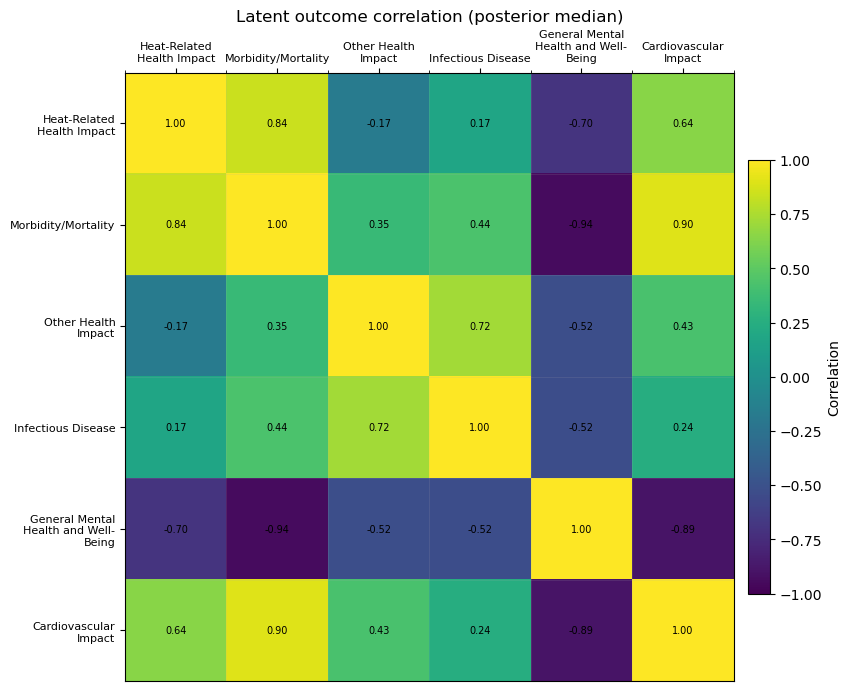}
    \caption{\textbf{Latent dependence among coarse health outcomes (posterior median).} Posterior median latent correlations between six coarse health-outcome indicators from a multivariate probit model. Values reflect residual dependence after accounting for shared covariates on the probit scale.}
    \label{fig:latent-corr-full}
\end{figure}

\paragraph{Outcome emphasis in the coded evidence base}
Among records with health-impact annotations ($n=12{,}065$), the distribution of labels is highly concentrated (Figure~\ref{fig:health-top40}). Temperature-related heat impacts are the single most common outcome tag (2{,}265 records; 18.8\% of health-coded records), followed by morbidity/mortality (1{,}289; 10.7\%), other health impacts (1{,}182; 9.8\%), infectious disease (1{,}115; 9.2\%), and general mental health and well-being (857; 7.1\%). Cardiovascular impacts and several infectious/vector-borne outcomes (e.g., dengue and malaria) also rank among the most frequently tagged outcomes, indicating that the coded literature emphasizes both acute temperature-related endpoints and infectious disease pathways.

\paragraph{Dyad concentration beyond marginal popularity}
Co-occurrence between exposure and health-impact terms departs strongly from independence and concentrates into a small set of highly enriched exposure--outcome dyads (Figure~\ref{fig:top-dyads}). The most extreme enrichments align with expected hazard--outcome mechanisms: \emph{Temperature: Extreme Heat/Heat Wave} and \emph{Temperature: Heat} co-occur with \emph{Heat-Related Health Impact} far more often than expected under independence (standardized residuals $|z| \gg 50$). Disaster-related exposures (floods and hurricanes) are strongly enriched with mental health outcomes (general mental health/well-being and stress/mood disorders), while \emph{Marine/Freshwater Biotoxin} is enriched with \emph{Marine Toxin Syndrome}. Respiratory allergy outcomes show enriched pairing with multiple allergen-related exposure tags (including air-pollution-related allergens and seasonality/temperature allergen tags), illustrating that the strongest dyads are not solely driven by the most common marginals but by specific, repeatedly studied exposure--outcome narratives.

\paragraph{Missingness is systematic: who gets health-outcome coding}
To assess whether health-impact annotation availability is plausibly missing-at-random, we modeled the probability that a record contains non-missing \texttt{health\_impact\_terms} as a function of publication year (centered), geography, and broad reference type. Figure~\ref{fig:missingness-logit-health} showed that annotation availability was strongly patterned by both time and geography. Using global/unspecified geography as the reference category, U.S.-focused records had higher odds of including health-impact terms (OR = 1.27, 95\% CI [1.17, 1.38]) and non-U.S.\ records have even higher odds (OR = 1.62, 95\% CI [1.52, 1.72]). Annotation availability also increased over time (OR per year = 1.02, 95\% CI [1.02, 1.03]). In contrast, reports were substantially less likely than journal articles to include health-impact term coding (OR = 0.44, 95\% CI [0.35, 0.57]), while the residual ``Other'' reference-type grouping was not distinguishable from journal articles in this model.

\paragraph{Residual dependence among coarse health outcomes}
Because multiple health-outcome labels can co-occur within the same record, we fit a multivariate probit model over six coarse outcome indicators and examined the posterior median latent correlation structure (Figure~\ref{fig:latent-corr}). Several outcome pairs showed strong positive residual correlation, including heat-related health impacts with morbidity/mortality ($\rho \approx 0.84$), morbidity/mortality with cardiovascular impacts ($\rho \approx 0.90$), and other health impacts with infectious disease ($\rho \approx 0.72$). In contrast, general mental health and well-being showed strong negative residual correlation with several other outcomes (e.g., morbidity/mortality $\rho \approx -0.94$; cardiovascular impacts $\rho \approx -0.89$; heat-related impacts $\rho \approx -0.70$), indicating that mental health tagging tended to occur in a distinct subset of records relative to these other coarse outcomes.

Across 45{,}423 exposure\(\times\)health cells, co-occurrence is strongly non-independent and concentrates in canonical hazard--outcome dyads (e.g., extreme heat/heat waves \(\leftrightarrow\) heat-related impacts; floods/hurricanes \(\leftrightarrow\) mental health and well-being; marine/freshwater biotoxins \(\leftrightarrow\) marine toxin syndromes; allergens \(\leftrightarrow\) upper respiratory allergy). Top cells are FDR-significant with very large standardized residuals and near-zero adjusted \(p\)-values, and a degree-offset bipartite Poisson edge model continues to identify many dyads as unusually strong beyond marginal term popularity.

A non-hierarchical asthma model was unstable, but hierarchical logistic regression with journal and geography random intercepts links asthma-tagged records primarily to air-pollution-related exposures (broad air pollution, ozone, particulate matter) and less to generic heat/temperature indicators (odds ratios \(<1\) for several heat-related terms). Methods and timescales shift over time: \texttt{model\_timescale} increases (OR \(\approx 1.18\) per year) while at least one legacy method tag declines (OR \(\approx 0.85\) per year). The declining legacy method tags are Methodology, Other Model/Methodology Type, Cost/Economic Impact Prediction, Exposure Change Prediction, Outcome Change Prediction, plus the Specify variants (SpecifyOther Model/Methodology Type, SpecifyCost/Economic Impact Prediction, SpecifyExposure Change Prediction, and Specify- discussion only). Coding completeness varies by time and geography---abstract availability rises with year, whereas exposure-term coding drops sharply in later years (Figure~\ref{fig:missingness}) and is predicted by geography---implying that naive prevalence trends can be confounded. LDA topic-share regressions show many strong associations with year, geography, and reference type, but topic labels/wordlists were not available in exported artifacts, limiting substantive topic naming.

\paragraph{Enrichment beyond marginal frequencies (Poisson-offset edge model)}
To separate genuinely enriched exposure--health pairings from those expected solely due to popular marginals, we compared observed co-occurrence counts to a Poisson offset baseline that adjusts for exposure and outcome frequencies. Figure~\ref{fig:poisson-offset-edges} shows that several dyads are markedly over-represented (large positive Pearson residuals), including \emph{Marine/Freshwater Biotoxin} $\rightarrow$ \emph{Marine Toxin Syndrome} and multiple allergen-related exposure tags pairing with allergic/respiratory and vector-/water-/food-borne outcomes. Conversely, several exposure--outcome combinations are strongly under-represented (large negative residuals), including broad climate and air-pollution exposure tags pairing with heat-related impacts more weakly than expected under the marginals, indicating that a portion of the co-occurrence structure is driven by specialization rather than general exposure popularity.

\paragraph{Geography-linked exposure profiles (hierarchical model)}
We next examined whether the mix of exposure tags differs systematically by geography using a geography-focused hierarchical logistic model with partial pooling. Figure~\ref{fig:geo-hier-ors} summarizes posterior odds ratios for exposure tags. The model indicates strong heterogeneity: broad \emph{Air Pollution} exposures show substantially elevated odds relative to the baseline (with posterior mass concentrated well above OR$=1$), while several temperature/disaster-related and context-specific exposures cluster closer to parity once uncertainty is accounted for. This pattern suggests that geographic emphasis in the coded literature is associated with a distinct exposure profile, and that hierarchical shrinkage is useful for stabilizing inference across many sparse exposure categories.

\subsection{Latent thematic trends from text}
\begin{figure}[ht]
    \centering
    \includegraphics[width=0.90\linewidth]{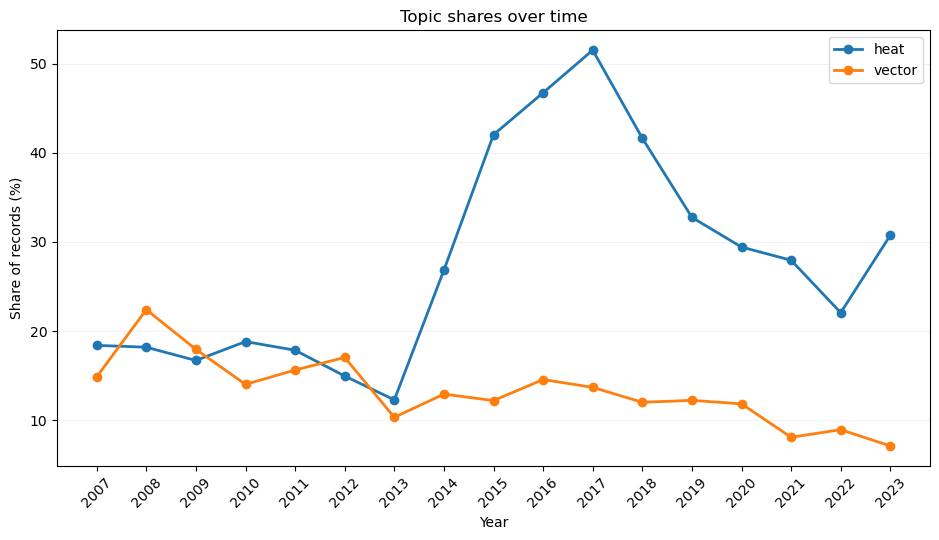}
    \caption{\textbf{Topic prevalence over time.} Annual share of records assigned to two illustrative latent themes (``heat'' and ``vector'') from LDA-style topic modeling of available text. Shares are computed within year among records with modelable text and should be interpreted as corpus-conditioned prevalence rather than population prevalence.}
    \label{fig:topic-shares}
\end{figure}

\paragraph{Shifting thematic emphasis over time}
To complement ontology-based tagging, we modeled latent themes from titles/abstracts using an LDA-style topic model and tracked annual topic prevalence. Figure~\ref{fig:topic-shares} contrasts two illustrative themes. The ``heat'' theme rises sharply in the mid-2010s (peaking around 2017) and remains elevated thereafter, whereas the ``vector'' theme shows a comparatively steady-to-declining trajectory over the same period. These trends provide a text-derived view of thematic emphasis that is consistent with a growing prominence of heat-related framing in the climate--health literature during the study window.

\section{Interpretations}
\subsection{Implications of heterogeneous coverage}
The completeness profile in Figure~\ref{fig:completeness_schema} implies that missingness in term fields should be interpreted primarily as \emph{unobserved annotation/extraction} rather than definitive absence of a concept in the underlying study. Consequently, prevalence estimates and exposure--health association patterns are conditional on the subset of records with available annotations, and temporal or geographic trends may reflect both substantive changes in the literature and shifts in annotation coverage. To mitigate misinterpretation, we report denominators for each analysis and restrict inference to within-subset comparisons when using sparsely available fields.

\subsection{Non-random annotation and multi-outcome structure}
Figure~\ref{fig:missingness-logit-health} indicates that health-outcome term availability is not missing-at-random: it varies strongly with geography, publication year, and reference type. This has two implications for inference. First, outcome prevalence estimates computed on annotated subsets should be interpreted as describing the \emph{coded} evidence base rather than the full corpus. Second, temporal or geographic trends in outcome tagging may partially reflect changing annotation coverage (e.g., increasing availability over time) rather than purely substantive shifts in the underlying literature. We therefore treat denominator reporting and subset-specific comparisons as essential when presenting results based on term fields.

Figure~\ref{fig:latent-corr} further suggests that the annotated literature is structured into partially distinct outcome bundles. Strong positive residual correlations (e.g., morbidity/mortality with cardiovascular outcomes; heat-related impacts with morbidity/mortality) are consistent with research clusters that jointly consider related endpoints, whereas the strong negative correlations involving mental health and well-being suggest a comparatively separate body of work (or differing annotation conventions) that emphasizes psychosocial outcomes rather than acute physiological endpoints. Practically, this dependence argues against treating outcome labels as independent in downstream analyses and motivates multi-outcome modeling when the goal is to characterize the structure of climate--health impacts rather than isolated endpoints.

\subsection{Growth is punctuated and the knowledge base is geographically and topically structured}
The \(\sim 10\%\)--\(11\%\) annual increase and multiple joinpoints suggest punctuated expansion rather than smooth exponential growth, consistent with waves of attention, funding cycles, and major climate/health events reshaping research priorities. The strong skew in U.S. geographic strata (\texttt{geo}) and non-U.S. grographic strata (\texttt{geo\_non\_us}) implies that aggregate topic and dyad summaries can be disproportionately driven by the majority-coded geography stratum, and that geography-aware modeling (e.g., random intercepts or stratified analyses) is necessary for credible comparisons across regions. Moreover, the substantial missingness in \texttt{geo\_non\_us} motivates treating it as a conditional attribute (relevant only for certain records) rather than a universally observed covariate.

\subsection{Core substantive signal: concentrated hazard--outcome narratives}
Figures~\ref{fig:health-top40} and \ref{fig:top-dyads} together suggest that the climate--health evidence base is not only expanding, but also structured around a relatively small number of repeatedly emphasized pathways. The dominance of heat-related impacts in the health-label distribution indicates that temperature-related endpoints provide a central organizing frame for the coded literature. At the same time, the strongest exposure--health dyads reflect coherent, mechanism-consistent narratives: extreme heat pairs tightly with heat-related outcomes, floods/hurricanes pair with mental health consequences, marine biotoxins pair with marine toxin syndromes, and allergen-related exposures pair with upper respiratory allergy.

Importantly, these findings should be interpreted as describing the \emph{annotated} evidence base rather than the full universe of climate--health research: dyad enrichment is computed on the subset of records with both exposure and health-impact coding, and outcome frequencies are computed on the subset with health-impact coding. Within those denominators, the strong enrichment patterns indicate that the taxonomy captures meaningful scientific structure (not merely popularity), while also highlighting potential path dependence in what combinations are repeatedly studied and labeled.

\subsection{Subfield specificity, and methodological evolution under missingness}
Log-linear screening and degree-offset network modeling converge on a core finding: a relatively small set of dyads (e.g., heat\(\leftrightarrow\)heat illness, floods/hurricanes\(\leftrightarrow\)mental health) are emphasized far beyond what would be expected from marginal frequencies alone. This likely reflects both scientific plausibility and path dependence in fundable/problem-framing conventions. Within respiratory research, after accounting for journal and geographic clustering, asthma-tagged studies align most strongly with air pollution, ozone, and particulate matter indicators, while heat/temperature terms are under-represented, underscoring that ``climate--health'' labeling spans subfields with distinct exposure emphases. The strong positive year association in ordered timescale models suggests increasing attention to projections/scenarios and longer planning horizons, while at least one legacy method indicator shows decline. Specifically, the legacy method tag SpecifyOther Model/Methodology Type declines over time. 

Finally, the finding that exposure-term presence declines over time---despite rising publication counts---is consistent with right-censoring in annotation; without explicit missingness modeling (or restriction to consistently coded windows), apparent shifts in topical prevalence may be artifacts of the curation pipeline rather than substantive changes in science. Because coding completeness varies by U.S. geographic strata (\texttt{geo}) in the missingness diagnostics (geography indicators predict the probability of having abstracts and exposure terms), naïve time-trend or prevalence analyses can be confounded by geography-dependent capture differences unless missingness is explicitly modeled or sensitivity-restricted to consistently coded subsets.

\subsection{Specialization, geographic stratification, and thematic shift}
The Poisson-offset residual analysis (Figure~\ref{fig:poisson-offset-edges}) indicates that the exposure--health association structure is not simply a byproduct of frequent marginals; instead, the literature contains strongly specialized dyads that occur far more often than expected under an independence baseline adjusted for exposure and outcome popularity. This supports the interpretation that the taxonomy captures meaningful, repeatedly studied mechanisms and narratives rather than arbitrary co-tagging.

The geography-focused hierarchical model (Figure~\ref{fig:geo-hier-ors}) suggests that geographic focus is associated with a distinct exposure mix, with some exposure categories (notably air-pollution-related tags) showing substantially elevated odds. Because estimates are partially pooled, the model also guards against over-interpreting rare exposures, emphasizing robust geographic stratification rather than noise from sparse categories.

Finally, the topic trend comparison (Figure~\ref{fig:topic-shares}) suggests a temporal shift in textual framing: heat-related thematic prevalence rises markedly from the mid-2010s onward, while vector-related themes do not increase in tandem. Together, these results point to an evidence base that is simultaneously (i) organized around specialized hazard--outcome pairings, (ii) stratified by geographic emphasis, and (iii) evolving in its dominant thematic narratives over time.

\section{Conclusion}
Using a multi-label bibliographic corpus of 22{,}695 climate--health records (2007--2023), we find: (i) rapid growth in publication volume with clear structural breaks, (ii) strong concentration of research attention around canonical exposure--health dyads that persists after adjusting for term popularity, (iii) robust alignment of asthma-tagged work with air-pollution exposures rather than temperature/heat, (iv) a pronounced shift toward longer modeling timescales with concurrent decline of at least one rare method tag, and (v) substantial, time-varying coding completeness---especially for exposure annotations---posing a key threat to naive temporal inference. Collectively, these results supported a view of the climate--health evidence base as simultaneously expanding and structurally concentrated, with annotation dynamics that must be modeled to make credible statements about changing topical emphasis over time.

\section*{Disclosure of LLM Use}
The authors used an LLM-based writing assistant to revise and improve portions of the manuscript prose based on text originally written by the authors. The authors retained full responsibility for the scientific content of the paper, including the research ideas, technical approach, experiments, results, and conclusions. All such revisions were carefully inspected by the authors for accuracy, and the manuscript was manually checked to ensure that no hallucinated, fabricated, or unsupported content was introduced.

\bibliographystyle{unsrt}
\bibliography{References}

\clearpage
\appendix
\section{Appendix Figures}

\setcounter{figure}{0}
\renewcommand{\thefigure}{A.\arabic{figure}}

\begin{center}
    \includegraphics[width=0.92\linewidth]{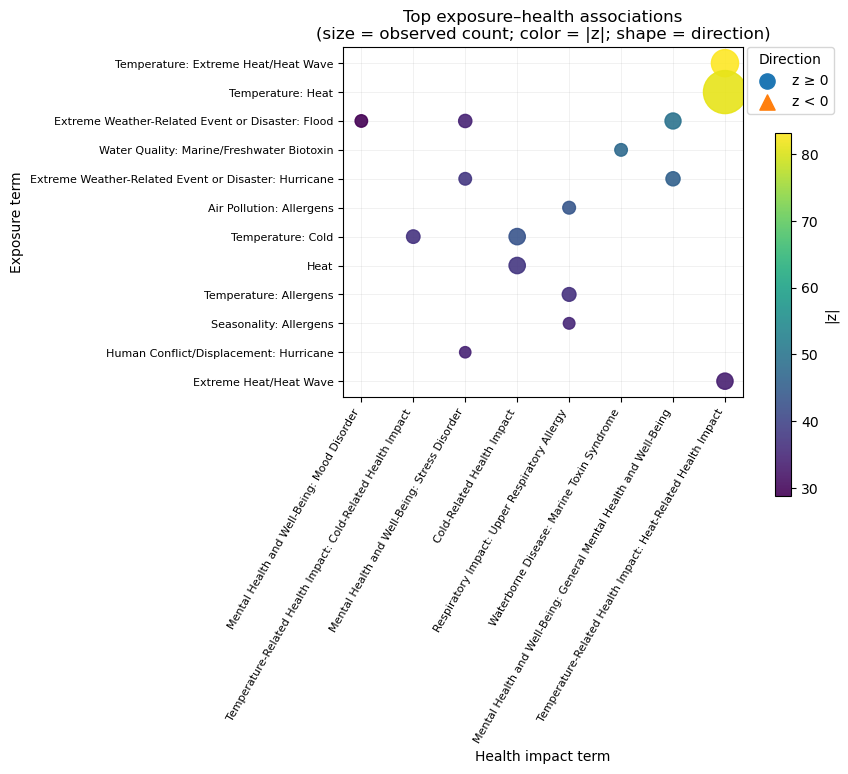}
    \captionof{figure}{\textbf{Top exposure--health associations under an independence baseline.} Bubble size denotes observed co-occurrence count; color denotes the magnitude of the standardized residual $|z|$ from an independence/log-linear baseline; marker shape denotes direction (over- vs.\ under-represented). Associations are computed on records with both exposure and health-impact annotations ($n=8{,}467$). Canonical dyads (e.g., extreme heat with heat-related impacts; floods/hurricanes with mental health outcomes; marine biotoxins with marine toxin syndromes; allergen-related exposures with upper respiratory allergy) are strongly over-represented.}
    \label{fig:top-dyads}
\end{center}

\begin{figure}[t]
    \centering
    \includegraphics[width=0.95\linewidth]{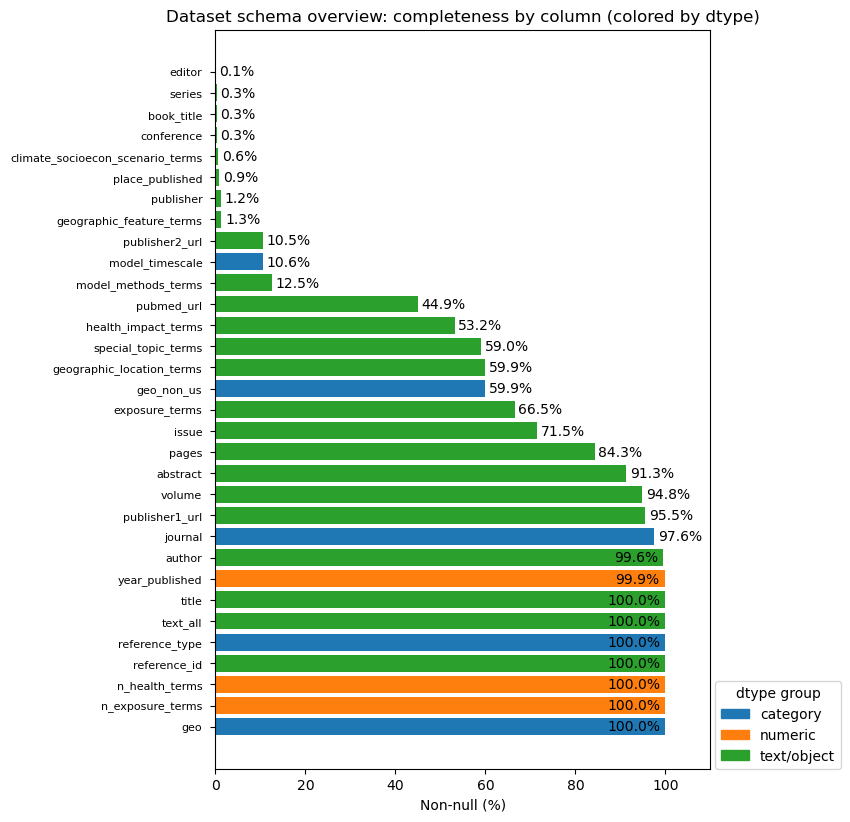}
    \caption{\textbf{Dataset schema and annotation coverage.} Column-wise non-null completeness across all records, colored by variable type (categorical, numeric, text/object). Core bibliographic fields are near-complete, while ontology-derived term fields (e.g., exposure and health impact terms) and modeling metadata (e.g., methods, timescale) are available for subsets of records, motivating denominator-aware analyses.}
    \label{fig:completeness_schema}
\end{figure}

\begin{figure}[ht]
    \centering
    \includegraphics[width=0.92\linewidth]{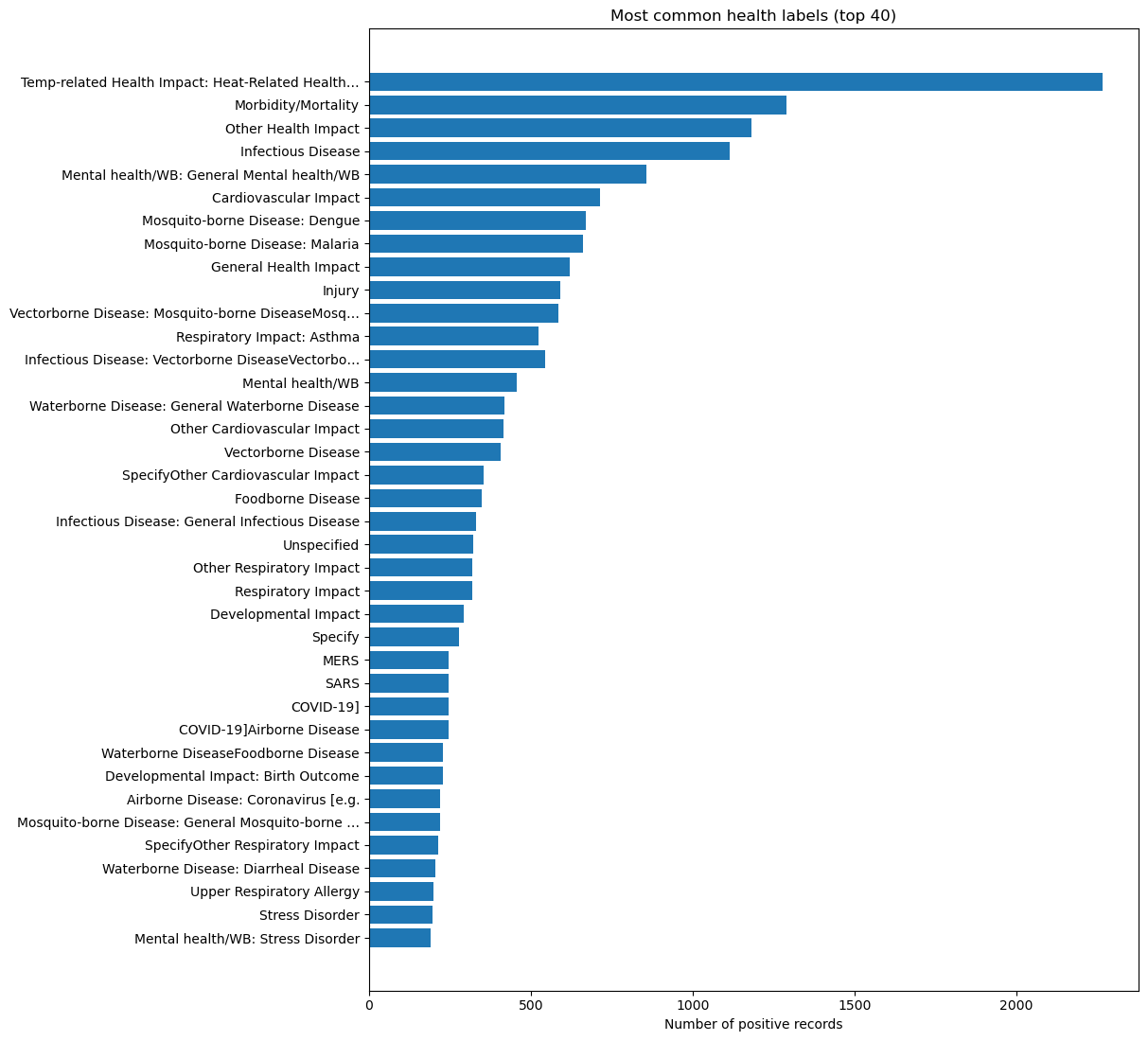}
    \caption{\textbf{Most common health-impact labels (top 40).} Counts reflect the number of records with each label present (binary presence per record) among records with non-missing health-impact annotations ($n=12{,}065$). Heat-related impacts dominate, followed by morbidity/mortality, infectious disease, mental health/well-being, and cardiovascular outcomes.}
    \label{fig:health-top40}
\end{figure}

\begin{figure}[ht]
    \centering
    \includegraphics[width=0.92\linewidth]{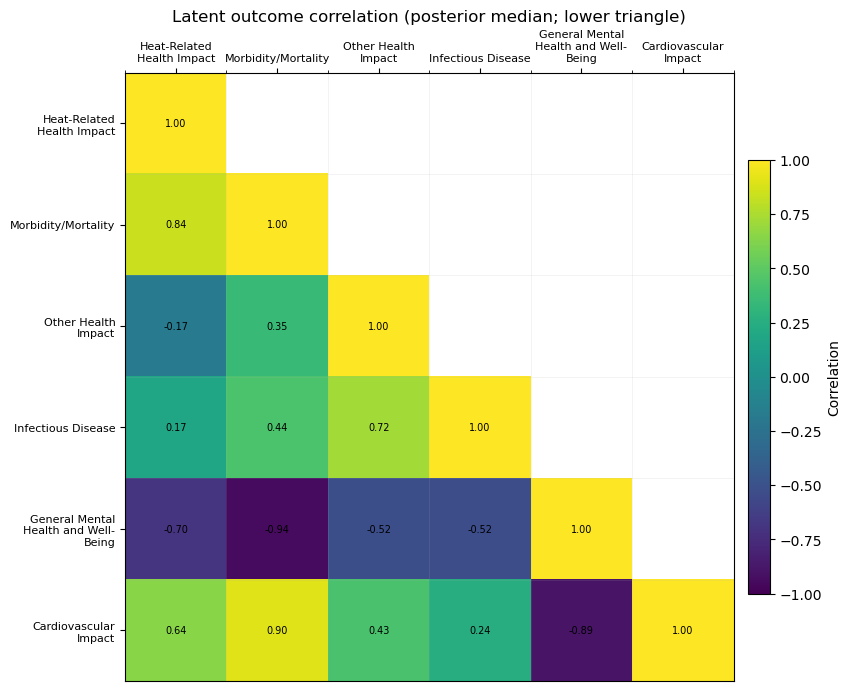}
    \caption{\textbf{Latent dependence among coarse health outcomes (posterior median; lower triangle).} Same estimates as Figure~\ref{fig:latent-corr-full}, shown as a lower-triangular matrix for readability.}
    \label{fig:latent-corr}
\end{figure}

\begin{figure}[ht]
    \centering
    \includegraphics[width=0.95\linewidth]{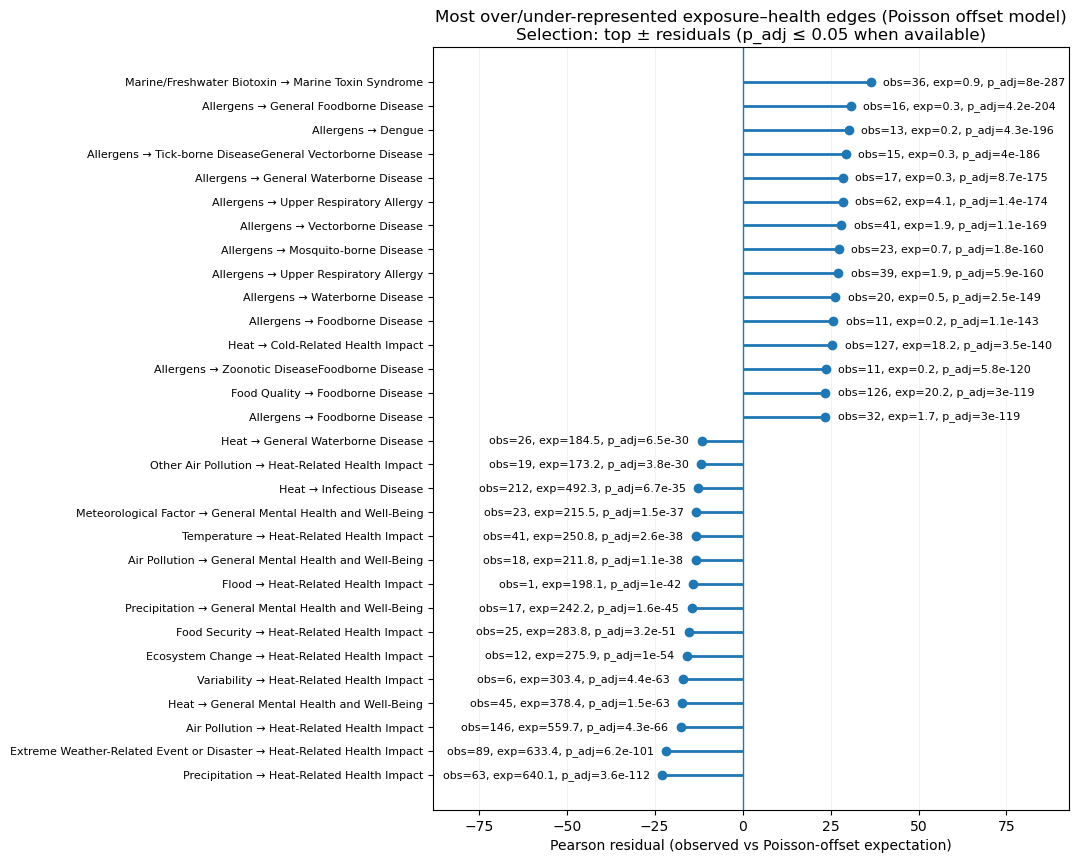}
    \caption{\textbf{Over/under-represented exposure--health edges under a Poisson offset baseline.} Pearson residuals compare observed co-occurrence counts for each exposure--health pair to expectations under a Poisson model with offsets that account for marginal exposure and outcome frequencies. Positive residuals indicate over-representation; negative residuals indicate under-representation. Labels report observed vs.\ expected counts and (when available) multiplicity-adjusted p-values. Computed on records with both exposure and health-impact annotations ($n=8{,}467$).}
    \label{fig:poisson-offset-edges}
\end{figure}

\begin{figure}[ht]
    \centering
    \includegraphics[width=0.95\linewidth]{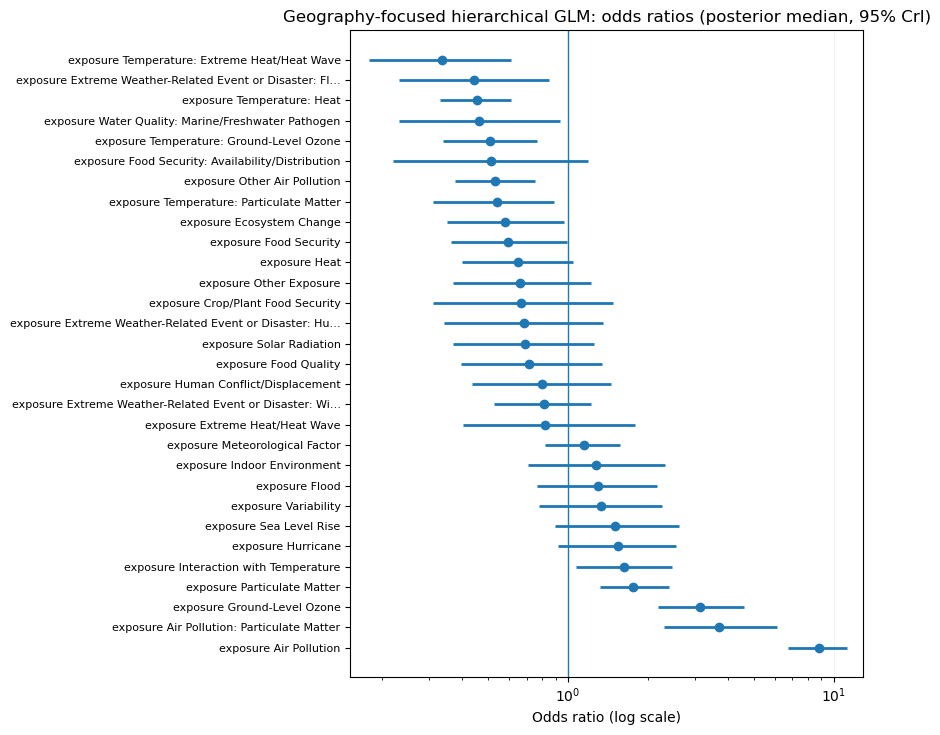}
    \caption{\textbf{Geography-focused hierarchical model: exposure odds ratios (posterior median, 95\% CrI).} Posterior odds ratios from a hierarchical logistic model assessing how exposure tags relate to geography-focused outcomes, with partial pooling to stabilize estimates across sparse exposures. Points show posterior medians and horizontal bars show 95\% credible intervals on a log-odds-ratio scale.}
    \label{fig:geo-hier-ors}
\end{figure}

\end{document}